\documentclass[onecolumn,showpacs,prd]{revtex4}
\usepackage{bm}
\usepackage{hyperref}
\usepackage{epsfig}

\newcommand{\NPA}[3]{Nucl.\ Phys.\ {\bf A#1}, #2 (#3)}

\newcommand{\PLB}[3]{Phys.\ Lett.\ B\ {\bf #1}, #2 (#3)}
\newcommand{\PRT}[3]{Phys.\ Rep.\ {\bf #1} #2 (#3)}
\newcommand{\PRL}[3]{Phys.\ Rev.\ Lett.\ {\bf #1}, #2 (#3)}

\newcommand{\PRC}[3]{Phys.\ Rev.\ C\ {\bf #1}, #2 (#3)}
\newcommand{\PRD}[3]{Phys.\ Rev.\ D\ {\bf #1}, #2 (#3)}
\newcommand{\JPG}[3]{J.\ Phys.\ G\ {\bf #1}, #2 (#3)}

\newcommand{\EL}[3]{Eur.\ Phys.\ Lett.\ {\bf #1},\ #2 (#3)}
\begin{document}

\title{Magnetized quark matter with a magnetic-field dependent coupling}

\author{Chang-Feng Li,$^{\mathrm{1}}$ Li Yang,$^{\mathrm{1}}$ Xin-Jian Wen,$^{\mathrm{1}}$\footnote{wenxj@sxu.edu.cn} and Guang-Xiong Peng $^{2,3}$\footnote{
gxpeng@gucas.ac.cn (G.X.Peng)} }

\affiliation{ $^1$Department of Physics and Institute of Theoretical
Physics, Shanxi University, Taiyuan 030006, China\\
$^2$College of Physics, University of Chinese
Academy of Sciences, Beijing 100049, China\\
$^3$ Synergetic Innovation Center for Quantum Effects and
Application, Hunan Normal University, Changsha, 410081, China
            }
\date{\today}
\begin{abstract}
It was recently derived that the QCD running coupling is a function
of the magnetic field strength under the strong magnetic field
approximation. Inspired by this progress and based on the
self-consistent solutions of gap equations, the properties of
two-flavor and three-flavor quark matter are studied in the
framework of the Nambu-Jona-Lasinio model with a magnetic-field
dependent running coupling. We find that the dynamical quark masses
as functions of the magnetic field strength are not monotonous in
the fully chirally broken phase. Furthermore, the stability of
magnetized quark matter with the running coupling is enhanced by
lowering the free energy per baryon, which is expected to be more
stable than that of the conventional constant coupling case. It is
concluded that the magnetized strange quark matter described by
running coupling can be absolutely stable.
\end{abstract}
 \maketitle
\section{INTRODUCTION}
The properties of strongly interacting quark matter under a strong
magnetic field have attracted much attention in the last decade
\cite{Miransky15}. The structure of dense matter and the behavior of
the interaction coupling constant will provide a new clue to the
comprehensive understanding of QCD theory under extreme conditions
\cite{Buballa05}. It has been proposed theoretically and
experimentally that a strong magnetic field could be present in the
core of neutron stars and in the noncentral collision experiments in
the Relativistic Heavy Ion Collider or the Large Hadron Collider
\cite{magnetic}. The typical strength of the strong magnetic fields
could be of the order of $10^{12}$ Gauss on the surface of pulsars.
Some magnetars can have even larger magnetic fields as high as
$10^{16}$ Gauss \cite{Thom96}. By comparing the magnetic and
gravitational energies, the physical upper limit to the total
neutron star is of order $10^{18}$ Gauss. And for the self-bound
quark stars, the limit could go higher \cite{Chanm01}. The maximum
strengths of $10^{18}\sim 10^{20}$ Gauss in the interior of stars
are proposed by an application of the viral theorem
\cite{magnetic,Dong91}. At the LHC/CERN energy, it is estimated to
produce a field as large as 5$\times 10^{19}$ Gauss
\cite{magnetic,Voronyuk}. It is thus reasonable to assume a uniform
and constant magnetic field to mimic the environment of the chiral
phase transition in heavy-ion collisions
\cite{allen13,Boomsma10,menezes09,Ferreira14}. The important effects
on the quark matter led by the strong magnetic field mainly include
the following two aspects. First, the strong magnetic field produces
the magnetic catalysis on the chiral phase transition at finite
temperature. Second, the charged fermions ruled by the Landau level
will display an anisotropic structure with respect to the direction
of the magnetic field. In fact, the behavior of quark matter is
mainly related to the quark condensate subject to the strong
magnetic field \cite{Ferreira14}. Consequently, the interaction
potential and the QCD ground state will be affected by the magnetic
field \cite{andrei13,Ferrer14}.

As is well known, the running behavior of the QCD coupling with
densities reflects the essential properties of strongly interacting
matter, which can be shown by solving the renormalization group (RG)
equation. In a strong magnetic field, the RG equation and the
polarization tensor will change due to the fact that charged
particles in a magnetic field obey the Pauli exclusion rule and the
Landau energy level arrangement \cite{Miransky,FerrerPRL11}.
Therefore, the magnetic-field-dependent coupling has been proposed
and verified recently \cite{Miransky,Ferreira14,wen2015}. Until now,
the effect of the magnetic-field dependence has been studied by
several versions of the analytic parametrization formula
$\alpha_s(eB)$ \cite{Miransky,Farias2014,Ferreira14}. The
investigation of the effect of the magnetic field on the coupling
constant can be summarized by two trends. One is to present an
analytic function of the running behavior at ultra-strong magnetic
field. The theoretical derivation of the magnetic-field-dependent
running coupling and the effective fermion mass in the propagator
can be obtained by the Schwinger-Dyson equation in the one-loop
approximation. The other is to fit the general parametrization
relation between the coupling constant and the magnetic field in
order to reproduce the critical temperature of the chiral symmetry
breaking from lattice QCD, since there is no direct result of the
running constant as a function of the magnetic field.

In the literature, the Nambu$-$Jona-Lasinio (NJL) model has been
widely employed in the study of the stability properties of strange
quark matter (SQM) without a magnetic field
\cite{Asakawa1989,Buballa1996,baldo07}. Recently, the NJL model has
been extended to the case of a strong magnetic field and many
special properties due to the magnetic field, for example, the
(inverse) magnetic catalysis effect \cite{catapaper}. It is
certainly expected that the stability of SQM is also strongly
affected by the magnetic field through the coupling constant. In
previous work using the NJL model, it was shown that a spherical
droplet of color-flavor locked (CFL) quark matter has a larger gap
energy when the coupling constant increases. A larger gap energy
will lead to lower free energy. Therefore, it is possible to find
absolutely stable CFL strangelets for a coupling constant $G$ larger
than some critical value \cite{Kiriyama}. However, the magnetic
field will rule out the constraint on the coupling constant
\cite{Klime05}. Namely, a droplet of magnetized quark matter may
exist for any value of $G$. In the quasiparticle model, we also
roughly found that for a proper value the magnetic field can enhance
the stability of the quark matter \cite{wen2013}. In the present
paper, we analyze the dynamical masses and the stability of quark
matter with the field-dependent running coupling in the NJL model.


The paper is organized as follows. In Sec. \ref{sec:model}, we
present the thermodynamics of the NJL model under a strong magnetic
field. The thermodynamical treatments in both SU(2) and SU(3)
versions are shown in the two subsections, respectively. In Sec.
\ref{sec:result}, the numerical results for the two-flavor and
three-flavor quark matter in $\beta$ equilibrium are presented, and
the discussions are focused on the stability properties and the
thermodynamical effect of the magnetic-field-dependent running
coupling. The last section is a short summary.

\section{Thermodynamics and stability of magnetized quark matter}
\label{sec:model} The dynamics of QCD matter are affected by strong
magnetic fields, especially with a magnitude of $eB\sim 15 m^2_\pi
~(\sim 10^{19}~\mathrm{Gauss})$ that can be produced in noncentral
relativistic heavy-ion collisions. In this paper, we mimic the
environment by assuming a uniform magnetic field in the $z$
direction, i.e., $A_\mu=\delta_{\mu 2}x_1 B$. First, we focus on the
two-flavor quark matter in the SU(2) NJL model. Then we continue to
study SQM in the SU(3) model.

\subsection{SU(2) NJL model in a strong magnetic field}
In the SU(2) version of the NJL model in a strong magnetic field,
the Lagrangian density reads
\begin{equation}{\cal{L}}_\mathrm{NJL}=\bar{\psi}( i/\kern-0.5em D-m )\psi
+G[(\bar{\psi} \psi)^2 +(\bar{\psi}i\gamma_5\vec{\tau} \psi)^2],
\end{equation}
where $\psi$ represents a flavor isodoublet ($u$ and $d$ quarks).
The coupling of the quarks to the electromagnetic field is
introduced by the covariant derivative
$D_\mu=\partial_\mu-iq_\mathrm{f} A_\mu$. Since the model is not
renormalizable at zero temperature, we should introduce a cutoff
$\Lambda$ in the 3-momentum space as in the usual way that has been
modified by a density-dependent momentum cutoff \cite{baldo07}.
Considering the general graphics of the dynamical fermion mass
generation in the Hartree (mean-field) approximation \cite{Ratti03},
the dynamical quark mass entering the thermodynamic potential at
finite chemical potential with a strong magnetic field is related to
the condensation term as

\begin{equation}M_i
=m_i-2G\langle\bar{\psi}\psi\rangle,\label{eq:gap}
\end{equation}where the condensation is $\langle\bar{\psi}\psi\rangle= \sum_{i=u,d}
\phi_{i}$ for the two-flavor case. The constituent mass of flavor
$i$ depends on both condensates. Therefore, we can always get the
same mass $M_u=M_d=M$, even for different charges and chemical
potentials. The contribution from the quark flavor $i$ is
\begin{equation}\phi_i=\phi_i^\mathrm{vac}+\phi_i^\mathrm{mag}+\phi_i^\mathrm{med}.\label{eq:condensate}
\end{equation}

The terms $\phi_i^\mathrm{vac}$, $\phi_i^\mathrm{mag}$, and
$\phi_i^\mathrm{med}$ representing the vacuum, magnetic field, and
medium contribution to the quark condensation are respectively
\cite{menezes09}
\begin{eqnarray}\phi_i^\mathrm{vac}&=&-\frac{M N_c}{2\pi^2} [\Lambda
\sqrt{\Lambda^2+M^2}-M^2\ln(\frac{\Lambda+\sqrt{\Lambda^2+M^2}}{M})],
\\
\phi_i^\mathrm{mag}&=& -\frac{M |q_i| B N_c}{2\pi^2} \left\{
\ln[\Gamma(x_i)]-\frac{1}{2}\ln(2\pi)+x_i-\frac{1}{2}(2x_i-1)\ln(x_i)
\right\},
\\
\phi_i^\mathrm{med}&=& \sum_{k_i=0}^{k_{i,\mathrm{max}}} a_{k_i}
\frac{M|q_i|BN_c}{2\pi^2} \ln\left[
\frac{\mu_i+\sqrt{\mu_i^2-s_i^2}}{s_i} \right].
\end{eqnarray}
The effective quantity $s_i=\sqrt{M^2+2k_i|q_i|B}$ sensitively
depends on the magnetic field. The dimensionless quantity is
$x_i=M^2/(2|q_i|B)$. The degeneracy label of the Landau energy level
is $a_{k_i}=2-\delta_{k0}$. The quark condensation is greatly
strengthened by the factor $|q_iB|$ together with the dimension
reduction $D-2$ \cite{Miransky,Kojo14}. The Landau quantum number
$k_i$ and its maximum $k_{i,\mathrm{max}}$ are defined as
\begin{equation}k_i\leq k_{i,\mathrm{max}}=
\mathrm{Int}[\frac{\mu_i^2-M^2}{2|q_i|B}],
\end{equation} where ``Int'' means the number before the decimal point.

The total thermodynamic potential density in the mean-field
approximation is
\begin{equation}\label{omega}\Omega=\frac{(M-m_0)^2}{4G}+\sum_{i=u,d}(\Omega_i^\mathrm{vac}+\Omega_i^\mathrm{mag}+\Omega_i^\mathrm{med}),
\end{equation}
where the first term in the summation is the vacuum contribution to
the thermodynamic potential, i.e.
\begin{eqnarray}\Omega_i^\mathrm{vac}= \frac{N_c}{8 \pi^2}
\left[M^4\ln(\frac{\Lambda+\epsilon_\Lambda}{M})-\epsilon_\Lambda
\Lambda (\Lambda^2+\epsilon_\Lambda^2) \right],
\end{eqnarray}where the quantity $\epsilon_\Lambda$ is defined as
$\epsilon_\Lambda=\sqrt{\Lambda^2+M^2}$. The ultraviolet divergence
in the vacuum part of the thermodynamic potential $\Omega$ is
removed by the momentum cutoff. In the literature, a form factor is
introduced in the diverging zero energy as a smooth regularization
procedure \cite{Gatto2010}. The magnetic field and medium
contributions are, respectively
\begin{eqnarray}
\Omega_i^\mathrm{mag}&=& -\frac{N_c (|q_i|B)^2}{2 \pi^2} \left[ \zeta'(-1,x_i)-\frac{1}{2}(x_i^2-x_i)\ln(x_i)+\frac{x_i^2}{4} \right], \\
\Omega_i^\mathrm{med}&=& -\frac{|q_i|B N_c}{4\pi^2}
\sum_{k=0}^{k_\mathrm{max}} a_{k_i} \left\{ \mu_i
\sqrt{\mu_i^2-(M^2+2k_i |q_i|B)}-(M^2+2k_i
|q_i|B)\ln[\frac{\mu_i+\sqrt{\mu_i^2-(M^2+2k_i
|q_i|B)}}{\sqrt{M^2+2k_i |q_i|B}}] \right\}\label{eq:ome_med},
\end{eqnarray} where
$\zeta(a,x)=\sum_{n=0}^\infty \frac{1}{(a+n)^x}$ is the Hurwitz zeta
function.
%
From the thermodynamic potential (\ref{omega}), one can easily
obtain the quark density as
\begin{eqnarray}
n_i(\mu,B)=\sum_{k=0}^{k_{i,\mathrm{max}}} a_{k_i} \frac{|q_i|
BN_c}{2\pi^2} \sqrt{\mu_i^2-(M^2+2k_i |q_i|B)}.
\end{eqnarray} The relevant pressure from the flavor $i$ contribution is
\begin{eqnarray}P_i(\mu_i,B)=-\Omega_i=-(\Omega_i^\mathrm{vac}+\Omega_i^\mathrm{mag}+\Omega_i^\mathrm{med}).
\end{eqnarray}
Under strong magnetic fields, the system's total pressure should be
a sum of the matter pressure and the field pressure contributions
\cite{menezes09,maxwell}. So we have
\begin{eqnarray}P_i(\mu_i,B)=-\Omega_i+\frac{B^2}{2},
\end{eqnarray}where the magnetic field term $B^2/2$ is due to the
electromagnetic Maxwell contribution. It is well known to us that
the energy density and pressure should vanish in vacuum. So the
pressure and the thermodynamic potential should be normalized by
requiring zero pressure at zero density as \cite{menezes09}
\begin{eqnarray}\label{eq:normali}P_i^\mathrm{eff}(\mu_i,B) = P_i(\mu_i,B)-P_i(0,B).
\end{eqnarray}In the normalization result, the field term is automatically absent.
According to the fundamental thermodynamic relation, the free energy
density at zero temperature is
\begin{eqnarray} \varepsilon_i =
-P_i^\mathrm{eff}+\mu_i n_i.
\end{eqnarray}

For asymmetric quark matter we should impose the $\beta$ equilibrium
by including the electron contribution under strong magnetic fields.
The electron chemical potential is not an independent variable and
can be expressed by the quark chemical potentials as
$\mu_e=\mu_d-\mu_u$. According to the similar normalization
procedure in Eq. (\ref{eq:normali}), it is required that
$P_{e,\mathrm{eff}}=P_e(\mu_e,B)-P_e(\mu_e,0)$. So the pressure of
electrons can be simplified as
$P_e^\mathrm{eff}=-\Omega_e^\mathrm{med}$ by setting $N_c=1$ and
$M=m_e$ in Eq. (\ref{eq:ome_med}). The corresponding number density
and the energy density are
\begin{eqnarray}n_e(\mu_e,B)&=&\sum_{k=0}^{k_{e,\mathrm{max}}} a_{k_i} \frac{|eB|}{2\pi^2}
\sqrt{\mu_e^2-(m_e^2+2k |eB|)},
\\
\varepsilon_e &=& -P_e^\mathrm{eff}+\mu_e n_e.
\end{eqnarray}
For the stellar matter in $\beta$ equilibrium, the charge neutrality
condition is
\begin{eqnarray}2n_u-n_d-3n_e=0.\label{eq:neutrality}
\end{eqnarray}
The system pressure and energy density are written as
\begin{eqnarray}P=\sum_i P_i^\mathrm{eff}, \ \ \ \ \
\varepsilon=\sum_i \varepsilon_i,
\end{eqnarray}where the summation goes over $u$, $d$ quarks and electrons.

The interaction coupling constant between quarks should in principle
be solved by the RG equation, or it can be phenomenologically
expressed in an effective potential \cite{Richard,Huangxg,JFXu15}.
In the infrared region at low energy, the dynamical gluon mass
represents the confinement feature of QCD \cite{Natale}.
Furthermore, in the presence of a strong magnetic field, the gluon
mass becomes large together with a decreasing of the interaction
constant, which leads to the damping of chiral condensation. For
sufficiently strong magnetic fields $eB\gg\Lambda^2_\mathrm{QCD}$,
the coupling constant $\alpha_s$ is proposed to be related to the
magnetic field as \cite{Miransky,Ferreira14}
\begin{eqnarray}\alpha_s(eB)=\frac{12\pi}{(11N_c-2N_f)\ln(|eB|/\Lambda_\mathrm{QCD}^2)}.
\end{eqnarray}
Motivated by the work of Miransky and Shovkovy \cite{Miransky}, a
similar ansatz of the magnetic-field-dependent coupling constant was
introduced in the SU(2) NJL \cite{Farias2014} and SU(3) NJL models
\cite{Ferreira14}. In the two-flavor version of the NJL model, based
on the lattice simulations, an interpolating formula was proposed as
\cite{Farias2014}
\begin{eqnarray}G'(eB)=\frac{G}{1+\alpha\ln(1+\beta|eB|/\Lambda^2_\mathrm{QCD})},
\label{eq:coupling}
\end{eqnarray}
where the energy scale is $\Lambda_\mathrm{QCD}=200$~MeV. The
parameters $\alpha=2$ and $\beta=0.000327$ are from the fit of the
lattice result for quarks condensates \cite{Farias2014}. We can find
that the running coupling constant versus the field $B$ gradually
approaches the constant value $G'(B\rightarrow 0)\sim G$.

\subsection{Magnetized strange quark matter in the SU(3) NJL model}

The SU(3) NJL Lagrangian density includes both a scalar-pseudoscalar
interaction and the 't Hooft six-fermion interaction[23] and can be
written as \cite{Grunfeld}
\begin{equation}{\cal{L}}_{NJL}=\bar{\psi}( i/\kern-0.5em D-m )\psi
+G\sum_{a=0}^8[(\bar{\psi}\lambda_a \psi)^2
+(\bar{\psi}\gamma_5\lambda_a
\psi)^2]-K\{\det_f[\bar{\psi}(1+\gamma_5)\psi]+\det_f[\bar{\psi}(1-\gamma_5)\psi]\}.
\end{equation}
The field $\psi=(u, d, s)^T$ represents a quark field with three
flavors. Correspondingly, $m=\mathrm{diag}(m_u, m_d, m_s)$ is the
current mass matrix with $m_u=m_d\neq m_s$. $\lambda_0=\sqrt{2/3}I$,
where $I$ is the unit matrix in the three-flavor space. $\lambda_a$
with $0<a\leq 8$ denotes the Gell-Mann matrix. Compared with the
two-flavor case, the gap equations for the three-flavor case are
coupled and should be solved consistently,
\begin{eqnarray}M_i-m_i+4G\phi_i-2K\phi_j\phi_k=0, \label{eq:gapSU3}
\end{eqnarray}
where $(i$, $j$, $k)$ is the permutation of $(u$, $d$, $s)$. The
quark condensates are the same as in Eq.(\ref{eq:condensate}).

The total thermodynamic potential density in the mean-field
approximation reads
\begin{equation}\label{omega}\Omega=2G \sum_{i=u,d,s}\phi^2_i -4K \phi_u\phi_d\phi_s
+\sum_{i=u,d,s}(\Omega_i^\mathrm{vac}+\Omega_i^\mathrm{mag}+\Omega_i^\mathrm{med}).
\end{equation}
The corresponding calculations of the normalized pressure and energy
density are similar to the SU(2) model.

The simple ansatz of the running coupling  is probably suitable for
the SU(3) NJL model if we include the $s$ quarks \cite{Ferreira14},
\begin{eqnarray}G'(eB)=\frac{G}{\ln(e+|eB|/\Lambda_\mathrm{QCD}^2)},\label{eq:G(eB)SU3}
\end{eqnarray} where the parameter $\Lambda_\mathrm{QCD}=300$ MeV,
which is different from the value in Eq.~(\ref{eq:coupling}).

\section{Numerical result and discussion}
\label{sec:result}

\subsection{Symmetric and asymmetric SU(2) quark matter}
\begin{figure}
\begin{center}
\includegraphics[width=0.40 \textwidth]{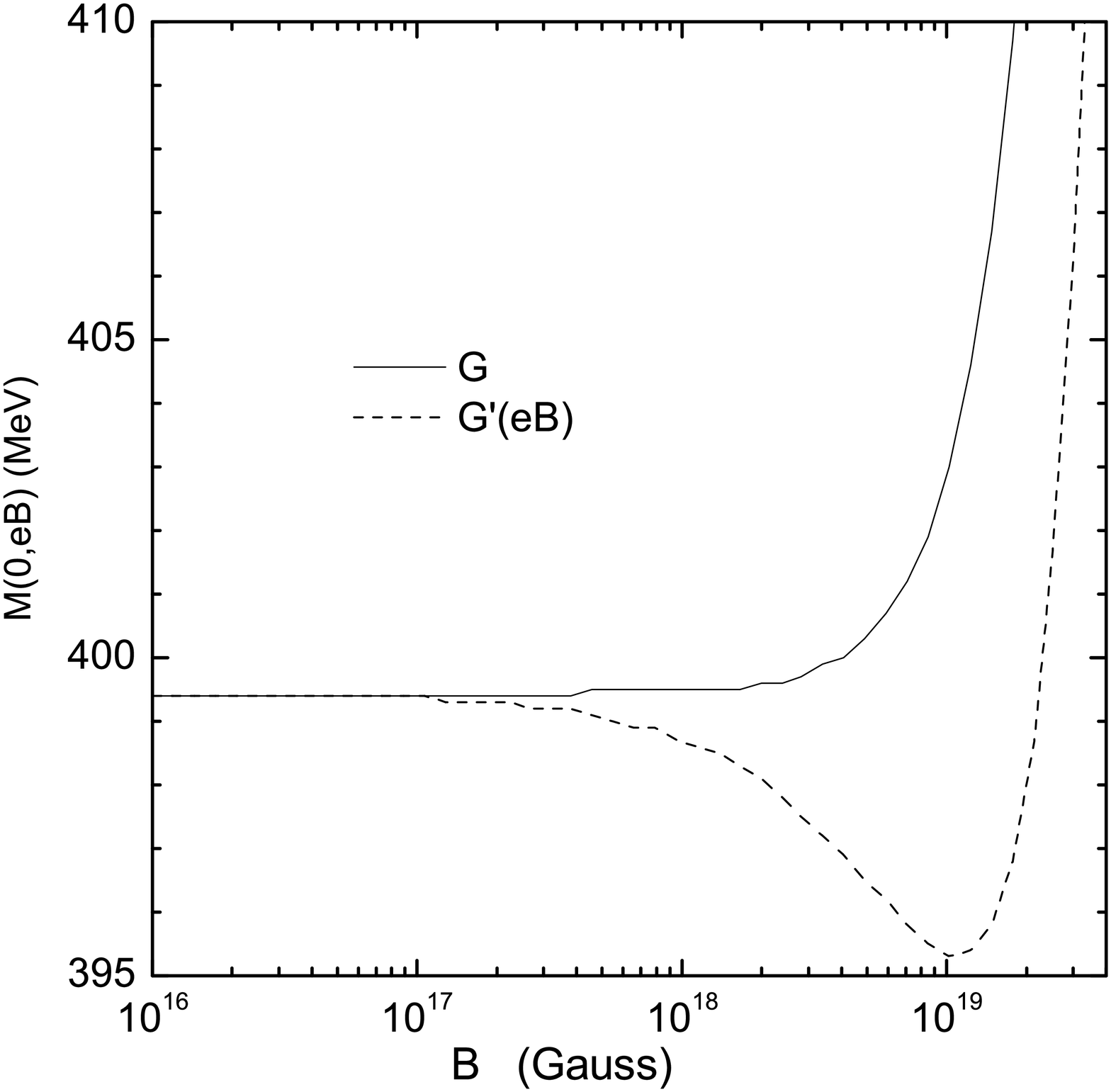}
\caption{\footnotesize Dynamical quark mass of SU(2) quark matter in
the fully chirally broken phase as a function of $B$ for the fixed
coupling constant $G$ and the running coupling $G'(eB)$.}
\label{Fig1}
\end{center}
\end{figure}
In the computation of this subsection, we consider the following set
of parameters for the SU(2) NJL model \cite{Hatsuda94}:
$\Lambda=587.9$ MeV, $N_c=3$, $m_u=m_d=5.6$ MeV, and
$G=2.44/\Lambda^2$. The corresponding electric charges are
$|q_d|=|q_s|=1/3=|q_u|/2$ in units of the elementary charge. First,
we investigate the symmetric quark matter with a common chemical
potential $\mu$ and common dynamical quark mass $M(\mu,eB)$ and do
the calculation under the coupling constant $G$ and the
magnetic-field-dependent running coupling $G'(eB)$ in Eq.
(\ref{eq:coupling}). The dynamical quark mass can be determined by
the gap equation (\ref{eq:gap}). It should be noticed that the gap
equation has more than one solution, with the physical being the one
that minimizes the thermodynamic potential. The zero chemical
potential case is the fully chirally broken phase. In Fig.
\ref{Fig1}, we show the dynamical quark mass $M(0,eB)$ as a function
of magnetic field strength $B$. The solid curve is for the fixed
coupling constant $G$. An increasing of the magnetic field leads to
an enhancement of the quark mass, which reflects the so-called
magnetic catalysis. The dashed curve is for the case of the running
coupling $G'(eB)$. It shows the distinct behavior of the effective
mass versus the magnetic field. In particular, by comparing the case
of the coupling constant, it is clear that the running coupling
constant $G'(eB)$ produces an inverse behavior of the dynamical mass
in the magnetic field range of $10^{17}\sim 10^{19}$~Gauss. In the
chiral-symmetry-broken phase, the quark effective mass will feel the
influence of the magnetic field through the correction to the quark
propagator. The numerical result in Fig.\ref{Fig1} shows that the
characteristic becomes more apparent for the magnetic field
$B=10^{19}$~Gauss, where the running coupling sensitively depends on
the magnetic field. But for smaller values of the magnetic field,
the two curves will gradually move closer to each other due to the
asymptotic behavior of the running coupling constant
$G'(eB\rightarrow0)\sim G$, where the coupling nearly remains
invariant.

\begin{figure}
\begin{center}
\includegraphics[width=0.40 \textwidth]{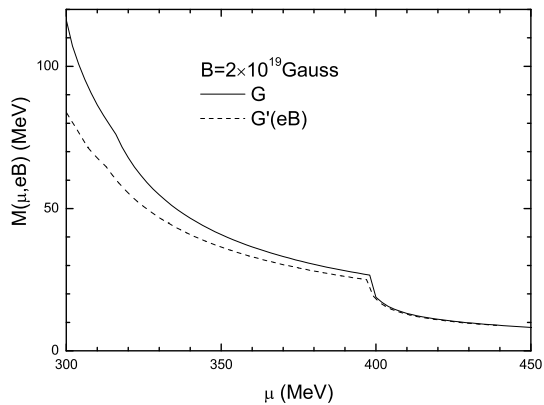}
\caption{\footnotesize Dynamical quark mass in the massive phase as
a monotonous decreasing function of the chemical potential for the
couplings $G$ and $G'(eB)$.} \label{Fig2}
\end{center}
\end{figure}

\begin{figure}
\begin{center}
\includegraphics[width=0.40 \textwidth]{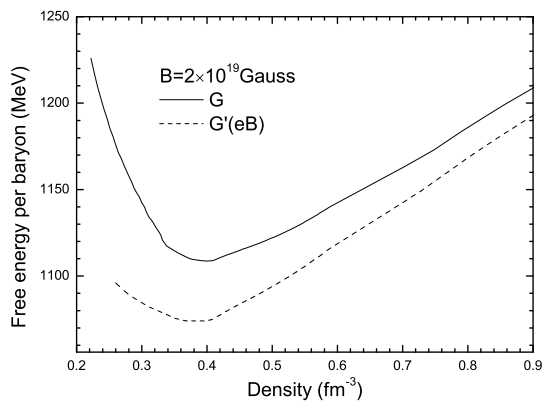}
\caption{\footnotesize The free energy per baryon of the symmetric
quark matter versus the baryon number density for the same parameter
sets as Fig. \ref{Fig2}.} \label{Fig3}
\end{center}
\end{figure}

For the massive phase with nonzero chemical potential at the
magnetic field $B=2\times 10^{19}$~Gauss, we can solve the gap
equation and calculate the dynamical quark mass $M(\mu,eB)$, which
is dependent on both the chemical potential and the given magnetic
field. In Fig. \ref{Fig2}, it is shown that the dynamical quark mass
will decrease as the chemical potential increases. The quark mass
under the running coupling $G'(eB)$ is lower than that of the fixed
coupling constant $G$ case, which is more clear in the small
chemical potential region. It shows that the correction of the
running coupling constant becomes very important near the infrared
region. Correspondingly, the free energy per baryon versus the
baryon number density is shown in Fig. \ref{Fig3}. The free energy
with the running coupling $G'(eB)$ is marked by the dashed curve,
which is lower than that of the fixed coupling case marked by the
solid curve. The value of the minimum of the free energy per baryon
on both curves is much bigger than the average energy value $930$
MeV for $^{56}$Fe. So it demonstrates that the two-flavor quark
matter is less stable than nuclear matter \cite{CJXia2015}, which is
in agreement with the Witten-Bodmer hypothesis \cite{Witten}.

\begin{figure}
\begin{center}
\includegraphics[width=0.40 \textwidth]{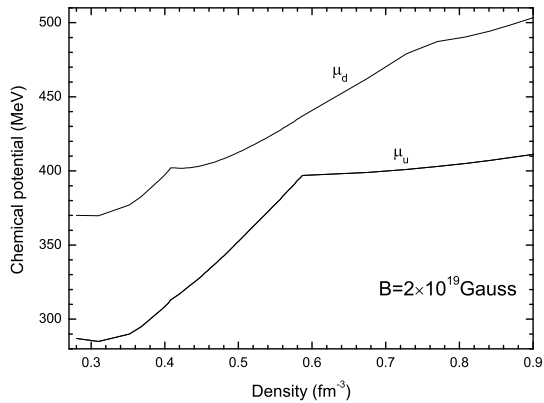}
\caption{\footnotesize The quark chemical potential versus the
baryon number density for the asymmetric quark matter at
$B=2\times10^{19}$~Gauss.} \label{Fig4}
\end{center}
\end{figure}

Now we study the isospin-symmetric quark matter by setting the
common chemical potential for $u$ and $d$ quarks. The isospin
symmetry can be broken under a strong magnetic field because of the
charge splitting for different flavors. We suppose that the quark
matter reaches the $\beta$ equilibrium condition. So the chemical
potentials satisfy $\mu_u+\mu_e=\mu_d$. The dynamic masses and the
two independent chemical potentials ($\mu_u$, $\mu_d$) can be
self-consistently solved by three equations: the gap equation
(\ref{eq:gap}), the baryon number conservation, and the charge
neutrality condition (\ref{eq:neutrality}). In Fig. \ref{Fig4} the
asymmetric chemical potentials are shown at the magnetic field
$B=2\times 10^{19}$~Gauss. The appearance of the inflection points
on the curves is due to the contribution of the Landau level. The
$d$ quark chemical potential $\mu_d$ is always much bigger than that
of the $u$ quark. In fact, it is naturally required that the number
of $d$ quarks is larger than the number of $u$ quarks in order to
reach global electrical neutrality together with the small amount of
leptons. On the other hand, it is understood that the Landau level
of the $d$ quark could be more than the $u$ quark level.

\begin{figure}
\begin{center}
\includegraphics[width=0.40 \textwidth]{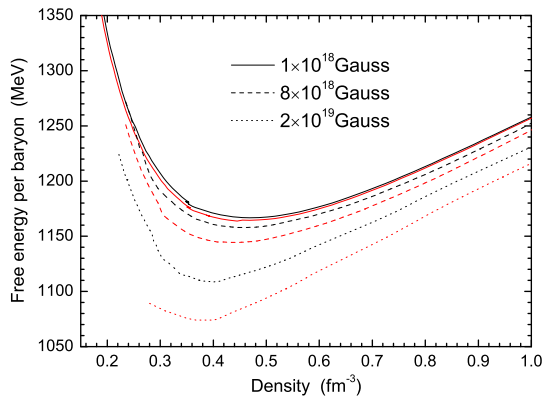}
\caption{\footnotesize The free energy per baryon of the asymmetric
two-flavor quark matter versus the baryon number density at the
three different magnetic field values. The three red curves are for
the running coupling $G'(eB)$.} \label{Fig5}
\end{center}
\end{figure}
In Fig. \ref{Fig5}, the free energy per baryon versus the baryon
number density is shown for different magnetic fields. The minimum
of the free energy per baryon is in the zero-pressure state. From
top to bottom, the magnetic fields of the three curves are,
respectively, $B=1\times 10^{18}$, $8\times 10^{18}$, and $2\times
10^{19}$~Gauss. It is known that the degeneracy factor of the quark
condensation is proportional to the magnetic field, so there will be
more quarks accommodated in the lowest Landau level for a larger
magnetic field. This is the reason why the quark matter will have
lower free energy at a stronger magnetic field. A larger magnetic
field will enhance the stability of quark matter by lowering the
free energy per baryon. Furthermore, we can find that the free
energy at the running coupling $G'(eB)$ (marked by red curves) is
all lower than that of the coupling constant $G$ at the same field
$B$.

\subsection{Numerical results of SU(3) NJL model}

It is necessary to extend the study of the stability of magnetized
quark matter to the SU(3) case. For the SU(3) NJL model, we adopt
the parameters $\Lambda=602.3$ MeV, $m_u=m_d=5.5$ MeV,
$m_s=140.7$~MeV, $G=1.835/\Lambda^2$, and $K=12.36/\Lambda^5$
\cite{Rehber96}. We assume that the three-flavor quark matter is in
$\beta$ equilibrium. Now there are three dynamical masses and two
independent chemical potentials, which can be determined by the
three gap equations (\ref{eq:gapSU3}), the baryon number
conservation, and the neutral charge condition,
\begin{eqnarray}2 n_u-n_d-n_s-3 n_e=0.
\end{eqnarray}

\begin{figure}
\begin{center}
\includegraphics[width=0.40 \textwidth]{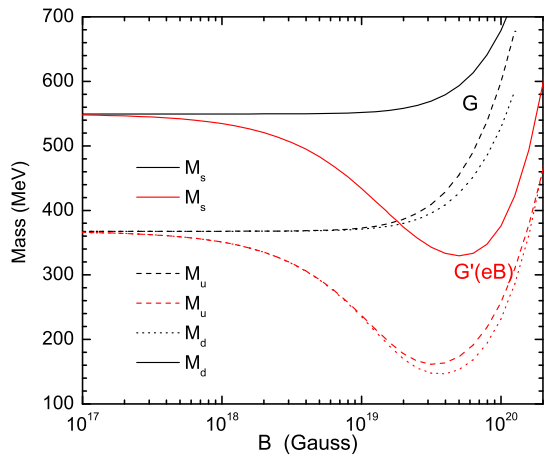}
\caption{\footnotesize Dynamical quark masses in the three-flavor
quark matter as functions of $B$ in the fully chirally broken phase.
The solid, dashed, and dotted curves from top to bottom denote the
masses $M_s$, $M_u$, and $M_d$ respectively. The red curves are for
the running coupling case.}\label{Fig6}
\end{center}
\end{figure}

In the fully chirally broken phase at zero chemical potential, the
dynamical quark masses only depend on the magnetic field. In Fig.
\ref{Fig6}, we show the dynamical quark masses of three flavors as
functions of the magnetic field. The dashed, dotted, and solid
curves are, respectively, for the $u$, $d$, and $s$ quarks. The
corresponding red ones represent the quark masses at the running
coupling $G'(eB)$ in Eq. (\ref{eq:G(eB)SU3}). As in the SU(2) case,
the running coupling produces different behavior for the dynamical
masses for all three flavors.

\begin{figure}
\begin{center}
\includegraphics[width=0.40 \textwidth]{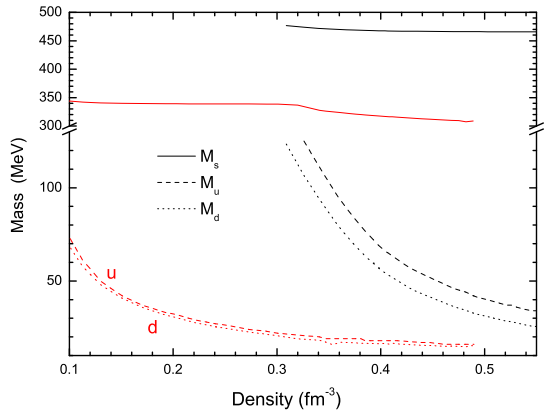}
\caption{\footnotesize Dynamical quark masses in the three-flavor
quark matter as functions of the baryon number density at the
magnetic field $B=2\times 10^{19}$~Gauss. The curves from top to
bottom denote $M_s$£¬ $M_u$, and $M_d$ respectively. The red curves
are for the running coupling case.} \label{Fig7}
\end{center}
\end{figure}

\begin{figure}
\begin{center}
\includegraphics[width=0.40 \textwidth]{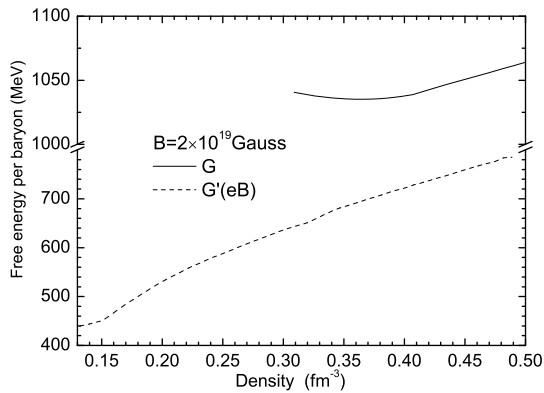}
\caption{\footnotesize The free energy per baryon of the
three-flavor quark matter versus the baryon number density at the
magnetic field strength $B=2\times 10^{19}$~Gauss.} \label{Fig8}
\end{center}
\end{figure}

We can solve the dynamical masses $M_i(\mu,eB)$ at finite chemical
potential in Fig.~\ref{Fig7}. The dynamical masses of $u$ and $d$
quarks apparently decrease as the density increases. At the coupling
constant $G=2\times 10^{19}$ Gauss, $M_s$ almost remains a constant
of $466$ MeV or so. Consequently, the strange quarks have no real
distribution in its Landau level in the system. On the contrary, the
running coupling $G'(eB)$ will lead to smaller dynamical masses
(marked by red curves). The strange quark mass $M_s$ decreases
slightly as the density increases, and thus the lowest Landau level
of the strange quark can appear at least. In Fig. \ref{Fig8}, we
compare the free energy per baryon under different coupling cases at
the same magnetic field $B=2\times 10^{19}$ Gauss. The upper solid
curve is for the coupling constant $G$ and the lower dashed curve is
for the running coupling $G'(eB)$. Therefore, it is possible that
the strange quark matter with a running coupling in a proper
magnetic field could be absolutely stable.

\section{summary}
We have studied the magnetized quark matter in the NJL model with a
magnetic-field-dependent running coupling to reflect the magnetic
field effect on the QCD vacuum structure and the interaction
potential between quarks. The effect becomes more important in the
infrared region. We studied the thermodynamic properties of both the
two-flavor and three-flavor quark matter in $\beta$ equilibrium
under a strong magnetic field.

In the NJL model, a magnetic field changes the quark dynamical mass
by modifying the quark condensation in the gap equations. In the
computation, we solved the gap equations for the fixed coupling $G$
and the magnetic-field-dependent running coupling $G'(eB)$,
respectively. First, in the fully chirally broken phase, we found
that the dynamical quark mass as only a function of the magnetic
field is not monotonous, contrary to the previous result for the
conventional fixed coupling constant. Furthermore, for two-flavor
quark matter in a larger magnetic field (about $10^{19}$ Gauss), the
running coupling $G'(eB)$ leads to a sharp drop of the dynamical
mass as the magnetic field increases, and a similar behavior appears
for three-flavor quark matter for a higher magnetic field than that
of the two-flavor case. Due to the reduction of the dynamical mass,
the strange quarks have a real distribution in the lowest Landau
energy level at least. Second, we found that the free energy per
baryon of the symmetric quark matter is smaller than that of the
asymmetric case, and it will decrease as the magnetic field strength
increases. Furthermore, we found that the stability of the
magnetized quark matter in $\beta$ equilibrium can be enhanced under
the running coupling by lowering the free energy. So the magnetized
SQM could be absolutely stable with the running coupling. In fact,
the comprehensive understanding of the QCD running coupling is
meaningful together with the one-loop vacuum and quark-gluon vertex
correction in the presence of a strong magnetic field \cite{Ayala},
which will greatly affect the chiral phase transition and the
stability properties of quark matter in a strong magnetic field. The
strong magnetic field will inevitably lead to the anisotropic
magnetization and pressure with respect to the direction of the
field \cite{Yang2011,Meneze15}. It is expected that the
field-dependent coupling would play a role in the anisotropic
structure and phase transition, which will be studied in the future.

\begin{acknowledgments}
The authors would like to thank support from the National Natural
Science Foundation of China (Nos. 11475110, 11135011, and 11575190)
and the Shanxi Provincial Natural Science Foundation under (Grant
No. 2013021006).

C.-F. L. and G.-X. P. contributed equally to this work.

\end{acknowledgments}

\end{document}